\documentclass[twoside,twocolumn,9pt]{article}
\usepackage{extsizes}
\usepackage[super,sort&compress,comma]{natbib} 
\usepackage[version=3]{mhchem}
\usepackage[left=1.5cm, right=1.5cm, top=1.785cm, bottom=2.0cm]{geometry}
\usepackage{balance}
\usepackage{times,mathptmx}
\usepackage{sectsty}
\usepackage{graphicx} 
\usepackage{lastpage}
\usepackage[format=plain,justification=raggedright,singlelinecheck=false,font={stretch=1.125,small,sf},labelfont=bf,labelsep=space]{caption}
\usepackage{float}
\usepackage{fancyhdr}
\usepackage{fnpos}
\usepackage[english]{babel}
\usepackage{array}
\usepackage{droidsans}
\usepackage{charter}
\usepackage[T1]{fontenc}
\usepackage[usenames,dvipsnames]{xcolor}
\usepackage{setspace}
\usepackage[compact]{titlesec}

\usepackage{amsmath}
\usepackage{amssymb}
\usepackage{hyperref}

\graphicspath{{./figures/}}

\definecolor{cream}{RGB}{222,217,201}

\begin{document}

\pagestyle{fancy}
\thispagestyle{plain}


\makeFNbottom
\makeatletter
\renewcommand\LARGE{\@setfontsize\LARGE{15pt}{17}}
\renewcommand\Large{\@setfontsize\Large{12pt}{14}}
\renewcommand\large{\@setfontsize\large{10pt}{12}}
\renewcommand\footnotesize{\@setfontsize\footnotesize{7pt}{10}}
\makeatother

\renewcommand{\thefootnote}{\fnsymbol{footnote}}
\renewcommand\footnoterule{\vspace*{1pt}%
\color{cream}\hrule width 3.5in height 0.4pt \color{black}\vspace*{5pt}} 
\setcounter{secnumdepth}{5}

\makeatletter 
\renewcommand\@biblabel[1]{#1}            
\renewcommand\@makefntext[1]%
{\noindent\makebox[0pt][r]{\@thefnmark\,}#1}
\makeatother 
\renewcommand{\figurename}{\small{Fig.}~}
\sectionfont{\sffamily\Large}
\subsectionfont{\normalsize}
\subsubsectionfont{\bf}
\setstretch{1.125}
\setlength{\skip\footins}{0.8cm}
\setlength{\footnotesep}{0.25cm}
\setlength{\jot}{10pt}
\titlespacing*{\section}{0pt}{4pt}{4pt}
\titlespacing*{\subsection}{0pt}{15pt}{1pt}

\fancyfoot{}
\setlength{\columnsep}{6.5mm}
\setlength\bibsep{1pt}

\makeatletter 
\newlength{\figrulesep} 
\setlength{\figrulesep}{0.5\textfloatsep} 

\newcommand{\topfigrule}{\vspace*{-1pt}%
\noindent{\color{cream}\rule[-\figrulesep]{\columnwidth}{1.5pt}} }

\newcommand{\botfigrule}{\vspace*{-2pt}%
\noindent{\color{cream}\rule[\figrulesep]{\columnwidth}{1.5pt}} }

\newcommand{\dblfigrule}{\vspace*{-1pt}%
\noindent{\color{cream}\rule[-\figrulesep]{\textwidth}{1.5pt}} }

\makeatother

\twocolumn[
\begin{@twocolumnfalse}
\vspace{3cm}
\sffamily
\begin{tabular}{m{4.5cm} p{13.5cm}}

 & \noindent\LARGE{\textbf{Decomposition mechanisms in metal\newline
borohydrides and their ammoniates}} \\
\vspace{0.3cm} & \vspace{0.3cm}\\

& \noindent\large{Evan Welchman\textit{$^{a}$} and Timo Thonhauser\textit{$^{a,b,c}$}}\\

 & \noindent\normalsize{
Ammoniation in metal borohydrides (MBs) with the form
$\mathcal{M}$(BH$_4$)$_x$ has been shown to lower their decomposition
temperatures  with $\mathcal{M}$ of low electronegativity ($\chi_p
\lesssim 1.6$), but raise it for high-$\chi_p$ MBs ($\chi_p \gtrsim
1.6$). Although this behavior is just as desired, an understanding of
the mechanisms that cause it is still lacking. Using \emph{ab initio}
methods, we elucidate those mechanisms and find that ammoniation always
causes thermodynamic destabilization, explaining the observed lower
decomposition temperatures for low-$\chi_p$ MBs. For high-$\chi_p$ MBs,
we find that ammoniation blocks B$_2$H$_6$ formation---the preferred
decomposition mechanism in these MBs---and thus kinetically stabilizes
those phases. The shift in decomposition pathway that causes the
distinct change from destabilization to stabilization around
$\chi_p=1.6$ thus coincides with the onset of B$_2$H$_6$ formation in
MBs.  Furthermore, with our analysis we are also able to explain why
these materials release either H$_2$ or NH$_3$ gas upon decomposition.
We find that NH$_3$ is much more strongly coordinated with
higher-$\chi_p$ metals and direct H$_2$ formation/release becomes more
favorable in these materials. Our findings are of importance for
unraveling the hydrogen release mechanisms in an important new and
promising class of hydrogen storage materials, allowing for a guided
tuning of their chemistry to further improve their properties.
} \\

\end{tabular}

\end{@twocolumnfalse} \vspace{0.6cm}]

\renewcommand*\rmdefault{bch}\normalfont\upshape
\rmfamily
\section*{}
\vspace{-1cm}

\footnotetext{\textit{$^{a}$Department of Physics, Wake Forest University, Winston-Salem, NC 27109, USA.}}
\footnotetext{\textit{$^{b}$Department of Chemistry, Massachusetts Institute of Technology, Cambridge, MA 02139, USA}}
\footnotetext{\textit{$^{c}$thonhauser@wfu.edu}}

\section{Introduction}\label{sec:introduction}

Hydrogen is an ideal energy carrier in a clean energy
system,\cite{Zuttel_2004:hydrogen_storage, Graetz_2009:new_approaches, Harrison_2015:materials_hydrogen}
but there is no known material to economically and safely store hydrogen
in a package sufficiently small and hydrogen-dense for use in mobile
applications.  In the search for a material that meets the US Department
of Energy's targets for a practical hydrogen storage
material,\cite{Basic_Research_Needs_2004, DOE_Targets_Onboard_2009}
metal borohydrides (MBs) have seen a surge of
interest\cite{Jain_2010:novel_hydrogen, Li_2011:recent_progress,
Rude_2011:tailoring_properties, Kostka_2007:diborane_release,
Yan_2012:pressure_temperature, Stadie_2014:supercritical_n2} and remain
one of the most promising classes of hydrogen storage
materials.\cite{Klebanoff_2013:5_years} In recent years,
various studies have found that these metal borohydrides typically
demonstrate improved hydrogen storage properties when complexed with
ammonia, forming a new class of materials called metal borohydride
ammoniates (MBAs).\cite{Jepsen_2014:boron-nitrogen_based,
Chu_2010:structure_hydrogen, Gu_2012:structure_decomposition,
Roedern_2015:ammine-stabilized_transition-metal, Sun_2012:new_ammine,
Yuan_2012:structure_hydrogen} Generally, MBAs tend to release hydrogen
gas at more practical temperatures and greater purity than their plain
MB counterparts.\cite{Jepsen_2014:boron-nitrogen_based}
Furthermore, recent experiments have mixed various MBAs with MBs or ammonia borane (NH$_3$BH$_3$) to tune the H$_2$ release temperature and purity.\cite{Huang_2015:metal-borohydride-modified_zrbh448nh3, Huang_2015:ammonia_borane}

\begin{figure}[t] \includegraphics[width=\columnwidth]{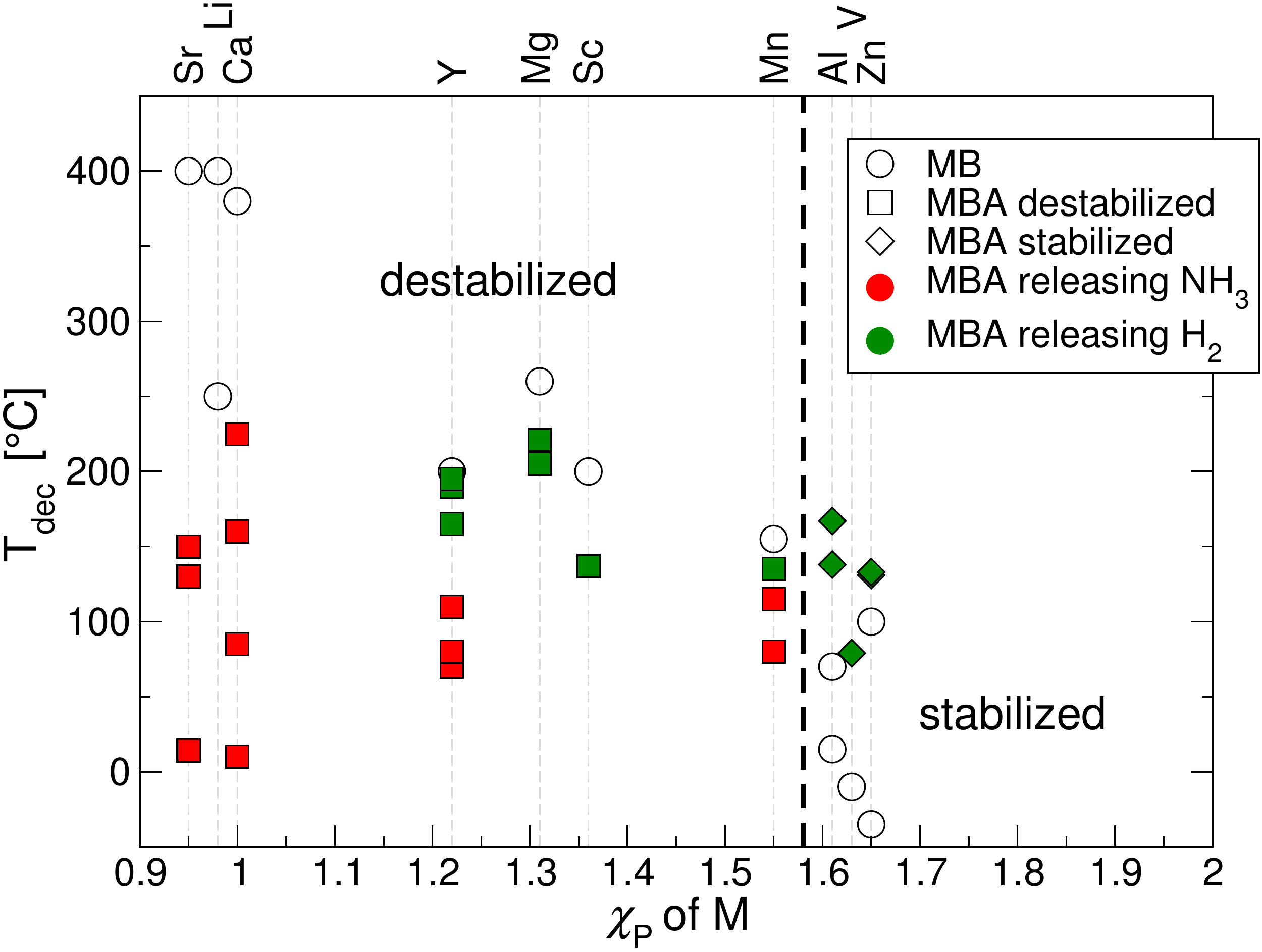}
\caption{\label{fig:stability}\small Experimentally observed
decomposition temperatures $T_\text{dec}$ for selected MBs and MBAs
plotted as a function of the electronegativity $\chi_p$ of the metal.
Red-filled symbols indicate that a material releases mostly NH$_3$,
while green-filled symbols indicate the release of mostly H$_2$.
Materials to the left of the dashed line (squares) are thermally
destabilized by ammoniation, to the right (diamonds) they are
stabilized. Data adapted from Jepsen et.\
al.\cite{Jepsen_2014:boron-nitrogen_based,
Jepsen_2015:trends_syntheses}} \end{figure}

Looking at the available experimental data for a large number of MBAs, a
clear pattern emerges between the Pauling electronegativity $\chi_p$ of
the metal $\mathcal{M}$ in $\mathcal{M}$(BH$_4$)$_x$ and the material's
decomposition temperature.\cite{Nakamori_2006:correlation_between} As
can be seen in Fig.~\ref{fig:stability}, ammoniating a MB lowers the
decomposition temperature of an MBA based on a metal with
electronegativity $\chi_p \lesssim 1.6$, but stabilizes an MBA based on
a high-electronegativity metal with $\chi_p \gtrsim 1.6$.  While this
experimental observation constitutes significant progress towards
improved hydrogen storage materials, the underlying mechanism causing it
is not understood. As such, further systematic improvements beyond
simple trial-and-error attempts are challenging. In this paper, we
present possible solutions and describe mechanisms that cause the
ammoniation to ``magically'' destabilize just those MBs that need it,
while stabilizing those that benefit from a higher decomposition
temperature.

Focusing on decomposition temperature alone, however, oversimplifies the
understanding of how these materials decompose. There are competing
decomposition mechanisms for each of these materials, resulting in
different gaseous products---mostly either NH$_3$ or H$_2$. These
products have been included in Fig.~\ref{fig:stability} and split the
destabilized set of materials into two distinct groups: those with low
electronegativity ($\chi_p \lesssim 1.2$) that release mostly NH$_3$ in
open systems and those with medium electronegativity ($ 1.2 \lesssim
\chi_p \lesssim 1.6$) that may release NH$_3$ at lower temperatures and
then transition to releasing mostly H$_2$ at higher temperatures.

In the decomposition pathway releasing ammonia, the thermodynamics are
such that releasing ammonia gas is more favorable than remaining in the
MBA structure.\cite{Jepsen_2015:trends_syntheses} As we will discuss in
detail below, our data indicates that the ammonia is more weakly bound
to the metal in MBAs where $\mathcal{M}$ has lower electronegativity. At
the temperature where this process occurs, the remaining MBA (or plain
MB) is below its decomposition temperature, and the MBA will not break
down further.  In high-electronegativity MBs ($\chi_p \gtrsim 1.6$),
ammoniation raises the material's decomposition temperature relative to
the plain MB and it was suggested that this behavior is linked to the
creation of a heteropolar dihydrogen bond network between the negatively
charged H$^{\delta-}$ in [BH$_4$]$^-$ and positively charged
H$^{\delta+}$ in NH$_3$.\cite{Soloveichik_2008:ammine_magnesium}
It is also known that the MBs in this category often decompose to
release diborane (B$_2$H$_6$),\cite{Harrison_2016:suppressing_diborane}
but this diborane is not released as a gaseous product in the presence
of NH$_3$.\cite{Jepsen_2015:ammine_calcium} Instead, it likely acts as
an intermediate, reacting with NH$_3$ to form boron nitrides as well as
H$_2$.\cite{Kostka_2007:diborane_release, Yan_2012:pressure_temperature,
Stadie_2014:supercritical_n2}

The decomposition of most MBs is a complicated process that is difficult
to model with \emph{ab initio} methods, as the correct intermediate
structures are often unknown. Jena et al.\ have tried to identify likely intermediate structures in the borohydride decomposition by simulating the structures of B$_n$H$_m$ clusters.\cite{Li_2010:reaction_intermediates} In the specific context of Mg borohydrides, Wolverton and co-workers have used a Monte Carlo-based structure-search method to find the most likely intermediate structures. \cite{Zhang_2012:theoretical_prediction} 
Another approach to explaining decomposition mechanisms in complex metal hydrides has been to simulate native defect formation in e.g.\ LiNH$_2$\cite{Hoang_2012:mechanisms_decomposition} and LiBH$_4$.\cite{Hoang_2012:mechanism_decomposition}
While these approaches have provided valuable insight and understanding of the decomposition 
processes in complex metal hydrides, the results may not directly apply to MBAs. 
A greater understanding of precisely how
and why these materials decompose will yield insight into how to design
an ideal hydrogen storage medium from this already attractive class of
materials.  

This work aims to elucidate various aspects of the
underlying mechanisms responsible for the decomposition processes in
MBAs. To this end, this manuscript is structured in the
follwing way: First, we examine the dihydrogen bond network of these
materials and find that it varies slightly with electronegativity, but
cannot explain the sharp distinction between destabilization and
stabilization observed in Fig.~\ref{fig:stability}. Next, we analyze the
thermodynamical stability and find that ammoniation always leads to a
lowering of stability. While that explains the lowering of materials
with $\chi_p \lesssim 1.6$ in Fig.~\ref{fig:stability}, it does not
explain the stabilization for $\chi_p \gtrsim 1.6$. Finally, directly
simulating the hydrogen release process, we find that those
high-$\chi_p$ materials get kinetically stabilized in that ammoniation
blocks the usual decomposition pathway.  The exact mechanism is material
dependent and we will explain it on a case-by-case basis.

\section{Computational Details}
\label{sec:comp_methods}

We performed calculations using density-functional theory (DFT) as
implemented in \textsc{vasp},\cite{Kresse_1996:efficient_iterative,
Kresse_1999:ultrasoft_pseudopotentials, Klimes_2011:van_waals} using the
included projector augmented wave (PAW) pseudopotentials and a plane wave
energy cutoff of 500~eV. We converged self-consistent energies to at
least 10$^{-5}$~eV and relaxed all atomic positions until all forces
dropped below 10~meV/\AA . Because of different unit cell sizes, k-point
meshes varied, but all used $\Gamma$ centered Monkhorst-Pack meshes
automatically generated in \textsc{vasp} with at least $20$$b$ k-points
in each direction, where $b$ is the length of the reciprocal lattice
vector in \AA$^{-1}$. We used the vdW-DF1
density-functional\cite{Thonhauser_2007:van_waals,
Langreth_2009:density_functional, Berland_2015:van_waals,
Thonhauser_2015:spin_signature} to properly account for the van der
Waals interactions in the dihydrogen bond networks in these materials.

We performed transition-state searches in \textsc{vasp} using the
climbing image nudged elastic band (NEB) method implemented in the
\textsc{vtst} package.\cite{Henkelman_2000:climbing_image,
Henkelman_2000:improved_tangent} In the case of B$_2$H$_6$ formation, we
generated end-point structures for these calculations by bringing
together nearby BH$_4$ molecules in the ground-state structures in a
directed perturbation of the system to form a stable B$_2$H$_6$ molecule
in the structure. This was done through an automated process that first
moved the BH$_4$ molecules away from the metal cation or cations to
which they were coordinated and then formed a B$_2$H$_6$ molecule along
the axis between the moved BH$_4$ molecules. The remaining two H atoms
were moved to the original locations of the BH$_4$ molecules. These
structures were then allowed to relax until all ionic forces dropped
below 10~meV/\AA.

In most cases, crystal structures are taken from experiment, but in some
cases we drew from theoretical work. Corresponding references are given
in Table~\ref{tab:structures}. For transition-state searches, we expanded any structures containing only one formula unit into a 2$\times$2$\times$2 supercell in order to avoid too much distortion in the crystal structure.

\begin{table}[t]
\caption{\label{tab:structures} Sources for the starting
structures of MBs and MBAs studied in this work.}
\begin{tabular*}{\columnwidth}{@{}l@{\hspace{0.5em}}l@{\extracolsep{\fill}}l@{\hspace{-1.5em}}r@{}}\hline
MB / MBA  & Ref. & MB/ MBA & Ref.     \\
\hline
Sr(BH$_4$)$_2\cdot$4NH$_3$       & \citenum{Jepsen_2015:ammine_calcium}                &
Mg(BH$_4$)$_2\cdot$2NH$_3$       & \citenum{Soloveichik_2008:ammine_magnesium}        \\

Sr(BH$_4$)$_2\cdot$2NH$_3$       & \citenum{Jepsen_2015:ammine_calcium}                &
Mg(BH$_4$)$_2$                   & \citenum{Harrison_2014:tuning_hydrogen}            \\

Sr(BH$_4$)$_2\cdot$NH$_3$        & \citenum{Jepsen_2015:ammine_calcium}                &
Zr(BH$_4$)$_4\cdot$8NH$_3$       & \citenum{Huang_2015:synthesis_structure}           \\

Sr(BH$_4$)$_2$                   & \citenum{Ravnsbk_2013:novel_alkali}                 &
Zr(BH$_4$)$_4$                   & \citenum{Rude_2012:synthesis_structural}           \\

Ca(BH$_4$)$_2\cdot$4NH$_3$       & \citenum{Jepsen_2015:ammine_calcium}                &
Mn(BH$_4$)$_2\cdot$3NH$_3$       & \citenum{Jepsen_2015:tailoring_properties}         \\

Ca(BH$_4$)$_2\cdot$2NH$_3$       & \citenum{Jepsen_2015:ammine_calcium}                &
Mn(BH$_4$)$_2\cdot$2NH$_3$       & \citenum{Jepsen_2015:tailoring_properties}         \\

Ca(BH$_4$)$_2\cdot$NH$_3$        & \citenum{Jepsen_2015:ammine_calcium}                &
Mn(BH$_4$)$_2$                   & \citenum{Harrison_2016:suppressing_diborane}       \\

Ca(BH$_4$)$_2$                   & \citenum{Majzoub_2009:crystal_structures}           &
Al(BH$_4$)$_3\cdot$6NH$_3$       & \citenum{Tang_2013:metal_cation-promoted}          \\

Y(BH$_4$)$_3\cdot$7NH$_3$        & \citenum{Jepsen_2015:trends_syntheses}              &
Al(BH$_4$)$_3$                   & \citenum{Harrison_2016:suppressing_diborane}       \\

Y(BH$_4$)$_3\cdot$6NH$_3$        & \citenum{Jepsen_2015:trends_syntheses}              &
Zn(BH$_4$)$_2\cdot$2NH$_3$       & \citenum{Gu_2012:structure_decomposition}          \\

Y(BH$_4$)$_3\cdot$5NH$_3$        & \citenum{Jepsen_2015:trends_syntheses}              &
Zn(BH$_4$)$_2$                   & \citenum{Harrison_2016:suppressing_diborane}       \\

Y(BH$_4$)$_3\cdot$4NH$_3$        & \citenum{Jepsen_2015:trends_syntheses}              &
NaZn(BH$_4$)$_3\cdot$2NH$_3$     & \citenum{Xia_2012:ammine_bimetallic}               \\

Y(BH$_4$)$_3\cdot$2NH$_3$        & \citenum{Jepsen_2015:trends_syntheses}              &
Li$_2$Mg(BH$_4$)$_4\cdot$6NH$_3$ & \citenum{Yang_2014:ammonia-stabilized_mixed-cation}\\

Y(BH$_4$)$_3\cdot$NH$_3$         & \citenum{Jepsen_2015:trends_syntheses}              &
Li$_2$Al(BH$_4$)$_5\cdot$6NH$_3$ & \citenum{Guo_2011:dehydrogenation_tuning}          \\

Y(BH$_4$)$_3$                    & \citenum{Sato_2008:experimental_computational}      &
                                 &                                                    \\
\end{tabular*}
\end{table}

\section{Results and Discussion}

\subsection{Examining the dihydrogen bond network}
\label{sec:dihydrogen}

A first attempt to explain the peculiar effect that ammoniation has on
the decomposition in Fig.~\ref{fig:stability} is to study the dihydrogen
bond networks in these materials.  All MBA materials contain networks of
heteropolar dihydrogen bonds (H$^{\delta -}
\cdot$$\cdot$$\cdot$H$^{\delta +}$) and it has been suggested that the
proximity of the protic N--H and hydridic B--H allows H$_2$ creation
through the simple combination of H$^{\delta +}$ and H$^{\delta
-}$.\cite{Soloveichik_2008:ammine_magnesium,
Jepsen_2014:boron-nitrogen_based} In low-$\chi_p$ materials, this
combination process is suspected to be a possible hydrogen production
mechanism,\cite{Jepsen_2015:tailoring_properties} and it was proposed
that stabilization of high-$\chi_p$ materials may be due to shielding of
the metal.\cite{Jepsen_2015:tailoring_properties,
Roedern_2015:ammine-stabilized_transition-metal} The dihydrogen bonds
clearly play some role in determining the stability of these materials,
as well as their decomposition, and further investigation is warranted.

\begin{figure}[t]
\includegraphics[width=\columnwidth]{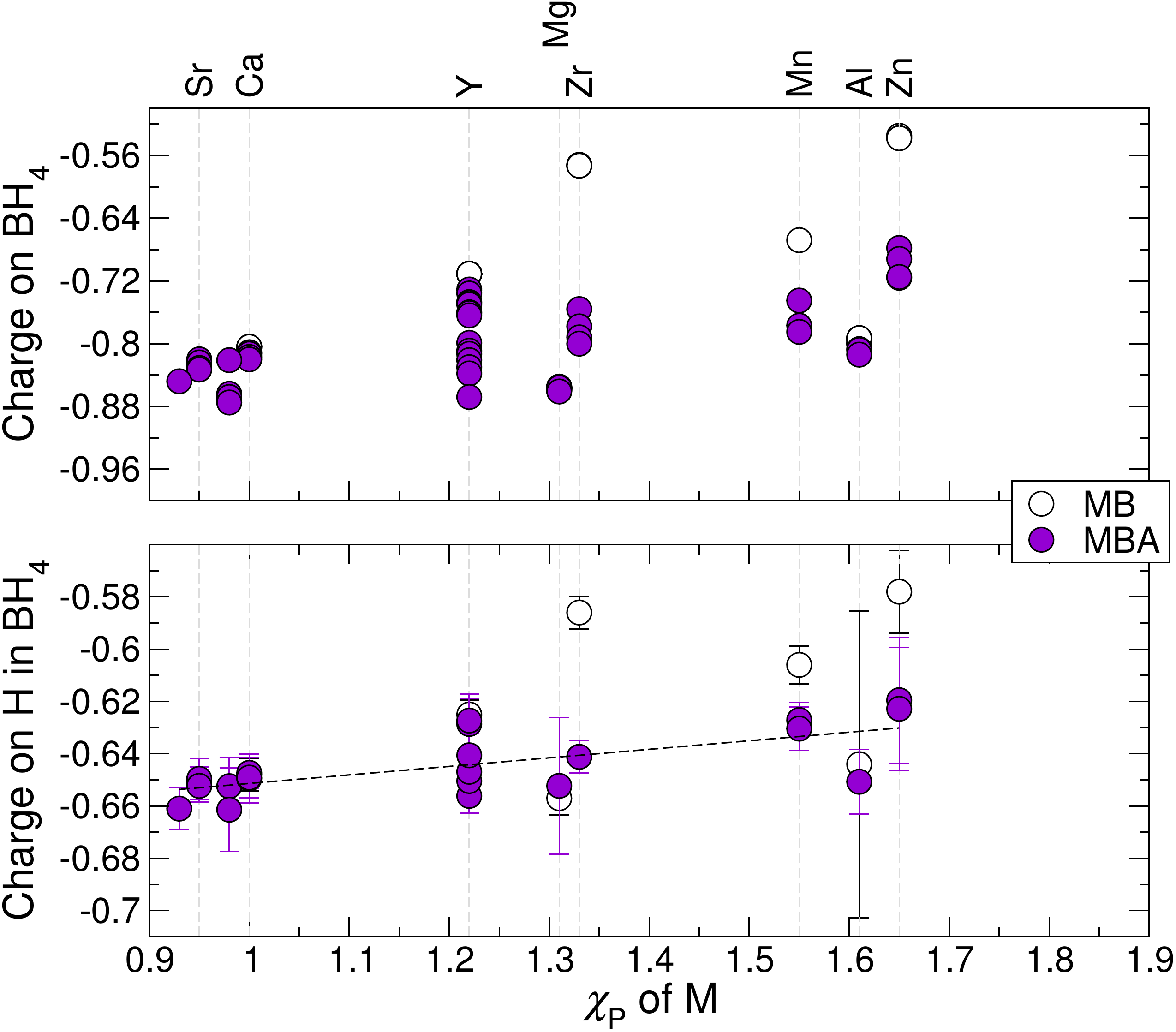}
\caption{\label{fig:BH4H_bader}(top) Bader charge [in units of $e$] on [BH$_4$]$^-$ anions relative to a neutral BH$_4$ molecule vs.\ $\chi_p$ of the metal cation with which they are coordinated. A perfect [BH$_4$]$^-$ anion would show a charge of $-1$. (bottom) Bader charge on  H$^{\delta -}$ in [BH$_4$]$^-$ anions relative to a neutral H atom. Error bars indicate one standard deviation from the mean. Dashed lines indicate linear fits for the mean values for MBAs.}
\end{figure}

\begin{figure}[t]
\includegraphics[width=\columnwidth]{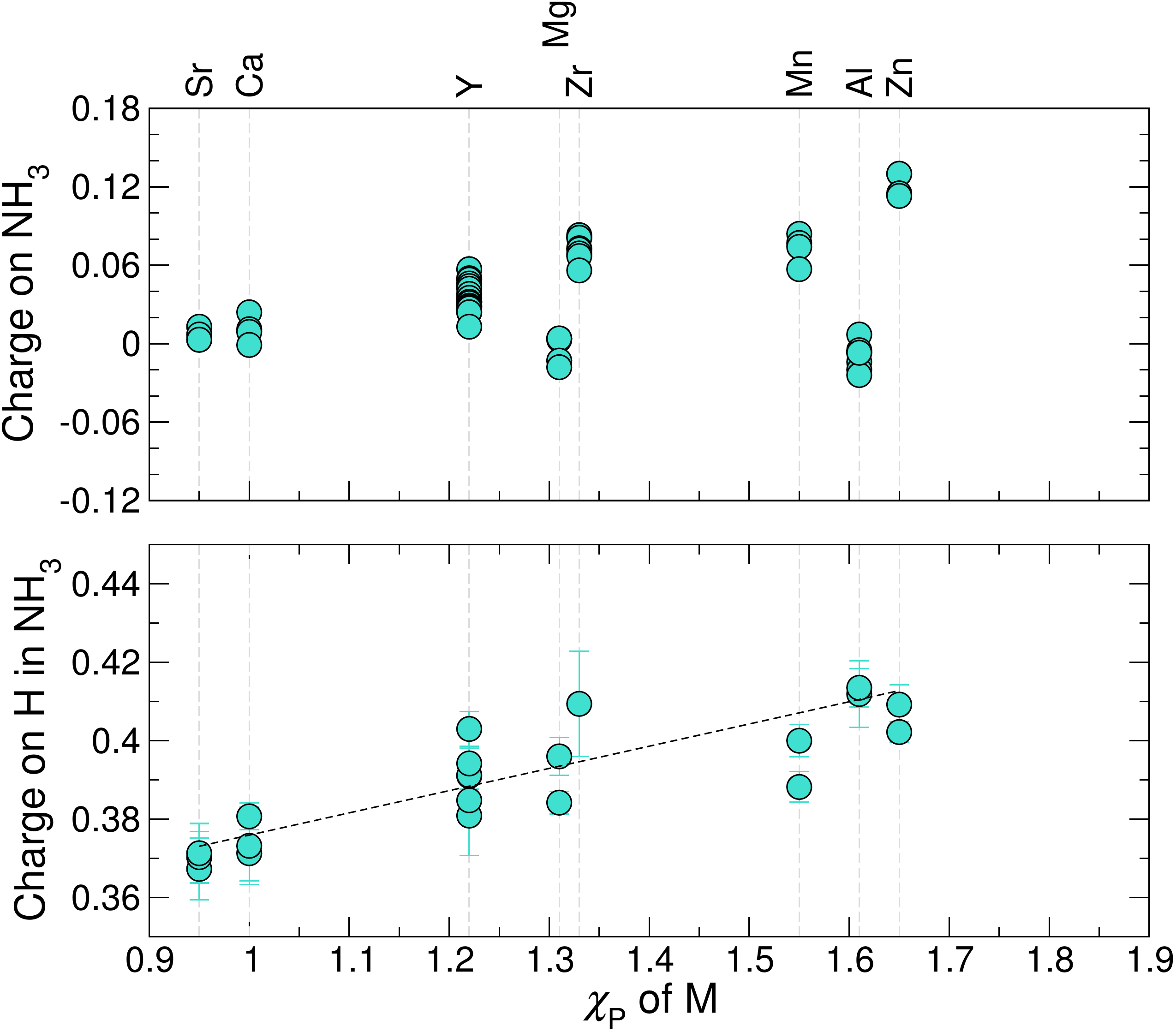}
\caption{\label{fig:NH3H_bader}(top) Bader charge [in units of $e$] on NH$_3$ molecules relative to a neutral NH$_3$ molecule vs.\ $\chi_p$ of the metal cation with which they are coordinated. (bottom) Bader charge relative to a neutral H atom for H$^{\delta +}$ in NH$_3$ molecules. Error bars indicate one standard deviation from the mean. Dashed lines indicate linear fits for the mean values for MBAs.}
\end{figure}

We start by analyzing these interactions through a charge partitioning
scheme and study how charge is distributed between the H atoms and the
small molecules in which they reside. Figures~\ref{fig:BH4H_bader} and
\ref{fig:NH3H_bader} show the Bader
charge\cite{Bader_1990:atoms_molecules, Henkelman_2006:fast_robust} of
BH$_4$ and NH$_3$ molecules in all MBAs from Table~\ref{tab:structures}
relative to neutral molecules of the same composition. The plots show
that the metal's electronegativity weakly affects the charge
distribution in these molecules. Higher electronegativity results in
weaker [BH$_4$]$^-$ anions and NH$_3$ molecules that begin to share
charge with other constituents of the material. 

On the level of individual H atoms, Fig.~\ref{fig:BH4H_bader} shows that
some materials demonstrate relatively large variations in the charge
held by H atoms in the same molecule. This variation is likely due to
how the molecules stack to form a crystal structure; the structure of
Al(BH$_4$)$_3$ is built from  Al(BH$_4$)$_3$ units where each BH$_4$
coordinates to only one Al atom, but the BH$_4$ units in Zn(BH$_4$)$_2$
act as bridges between multiple Zn atoms. Consequently, the charge on a
BH$_4$ anion in Al(BH$_4$)$_3$ is more concentrated on one side of the
molecule, giving the H atoms in Al(BH$_4$)$_3$ a larger variation in
charge than those in Zn(BH$_4$)$_2$.

\begin{figure}[t]
\includegraphics[width=\columnwidth]{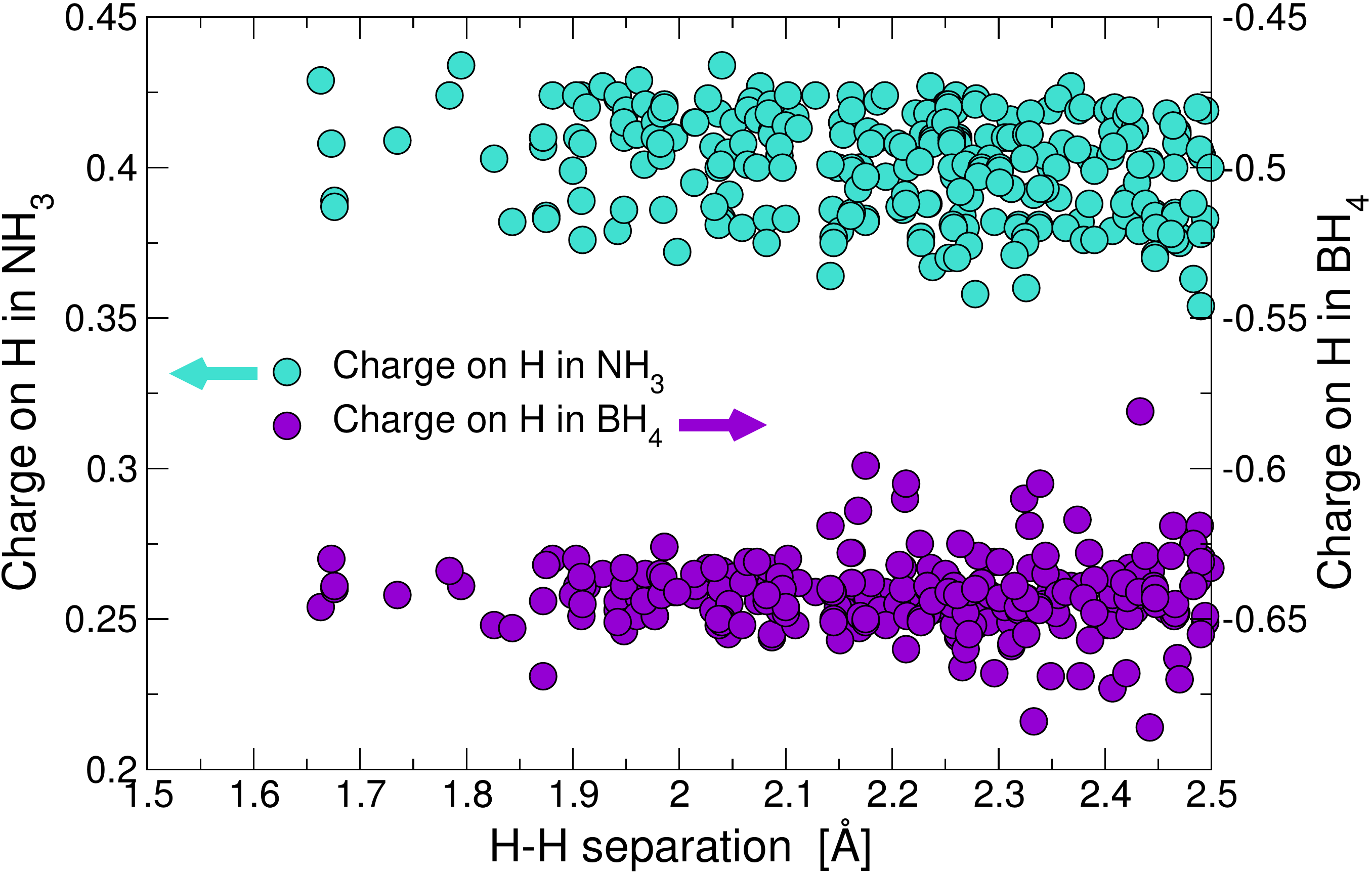}
\caption{\label{fig:dihydrogen_q} Bader charge [in unites of $e$] on H atoms in NH$_3$ and BH$_4$ molecules that are part of heteropolar dihydrogen bonds as a function of the H--H separation. All MBAs from Table~\ref{tab:structures} are considered.}
\end{figure}

\begin{figure}[t]
\includegraphics[width=\columnwidth]{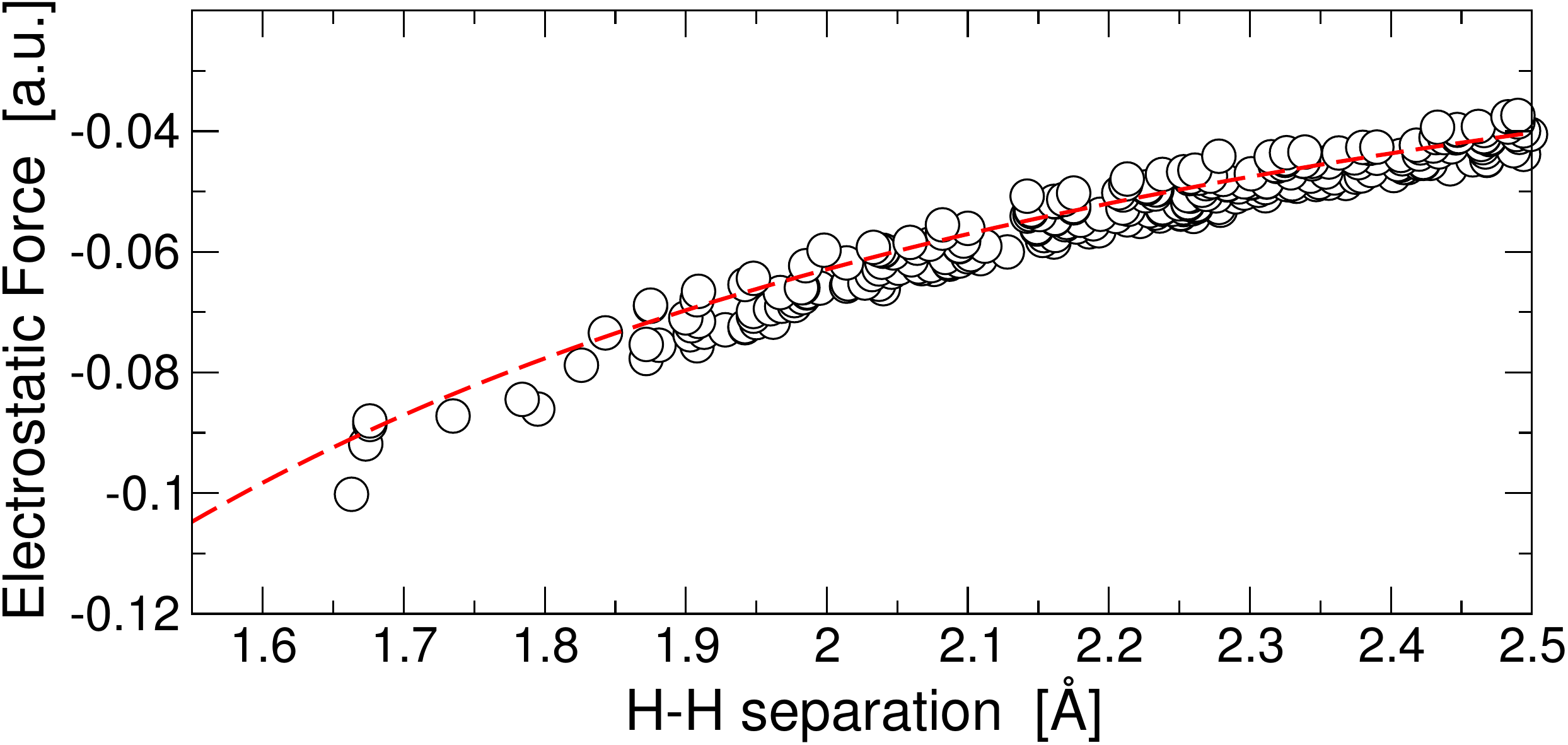}
\caption{\label{fig:dihydrogen} Modeled electrostatic force between H$^{\delta +}$ and H$^{\delta -}$ of neighboring molecules across all MBAs studied. The dashed red line
shows the exact Coulomb interaction calculated from the average H$^{\delta +}$ and H$^{\delta -}$ charges from all MBAs in Table~\ref{tab:structures}. Deviations from the exact interaction are very
small.}
\end{figure}

We also investigated the effect of H$\cdot$$\cdot$$\cdot$H bond length
on how much charge builds up on the H atoms in these materials; one
might expect that shorter bond lengths would draw stronger relative
charges for H$^{\delta +}$ and H$^{\delta -}$.
Figure~\ref{fig:dihydrogen_q} shows that H$^{\delta +}$ in NH$_3$ on
average does draw more charge when involved in shorter dihydrogen bonds,
but the relationship is weak and H$^{\delta -}$ shows no relationship at
all. To demonstrate the strength of this relationship, we modeled the
dihydrogen bond as a Coulomb interaction ($q_1q_2 / r^2$ with calculated Bader charges of H$^{\delta +}$ and H$^{\delta -}$ as $q_1$ and $q_2$). Figure~\ref{fig:dihydrogen}
shows that the data points fall almost perfectly along a fit
generated by using the average H$^{\delta +}$ and H$^{\delta -}$ charges
that we calculated from all MBAs included in Table~\ref{tab:structures}. Thus, bond length has a weak effect on charge buildup on
H$^{\delta +}$ atoms, and the corresponding increase in the
electrostatic force across the bond is very small. This means that there
is almost no additional charge polarization across the NH$_3$ and
[BH$_4$]$^-$ molecules due to short dihydrogen bond lengths and any
significant effects are localized to the electron clouds around the H
atoms.

To summarize, $\chi_p$ and the specific geometry of the structure are
more important for determining the charge distribution in the material
than bond lengths.  These results indicate that the dihydrogen bond
network is very similar between different MBAs.  While the dihydrogen
bond network may play an important role in the decomposition process,
that role does not appear to be linked in an obvious way with $\chi_p$.
It follows that variations (by themselves) in the dihydrogen bond
network from one MBA to another cannot explain the distinct
destabilizing/stabilizing effect of ammoniation observable in
Fig.~\ref{fig:stability}.

\subsection{Examining thermodynamic stability}

\begin{figure}[t]
\includegraphics[width=\columnwidth]{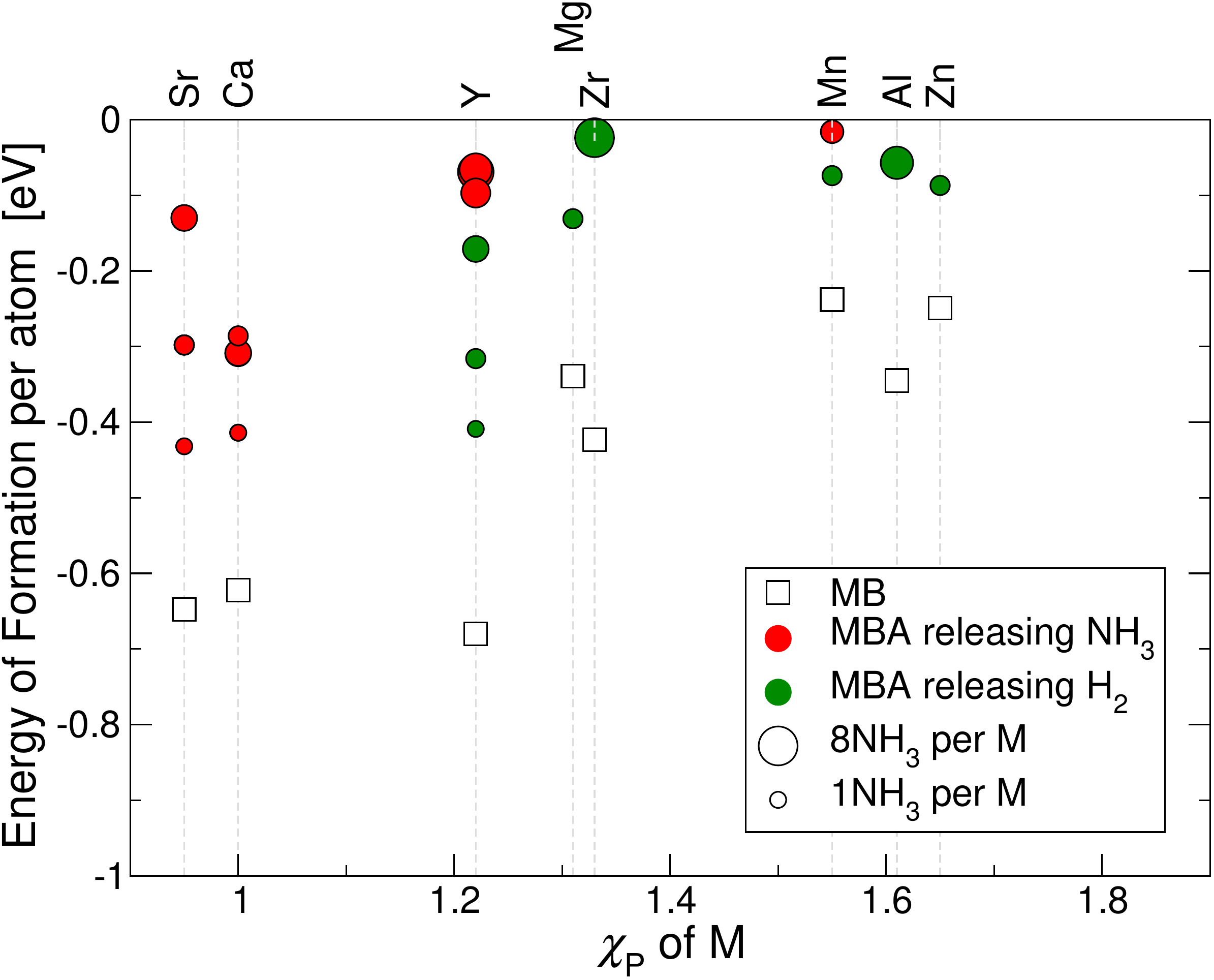}
\caption{\label{fig:Energy_of_form} Energy of formation per atom for pure MBs and MBAs relative to the constituent elements in their natural states. Red and green spots indicate MBAs that release NH$_3$ and H$_2$ upon decomposition, respectively. The size of the symbol indicates the number of NH$_3$ molecules in the material per metal. In all cases studied, the ammoniated MB is destabilized by the addition of the NH$_3$. See also Table~\ref{tab:formation_energy}.}
\end{figure}

Moving beyond the dihydrogen bond network, we next examine the
thermodynamics of the ammoniation process by calculating the energy of
formation for both MB and MBA materials relative to their constituents
in their natural states (i.e.\ the appropriate, separate amounts of solid $\mathcal{M}$ and  B, and gaseous N$_2$ and H$_2$). 
The results are given in
Table~\ref{tab:formation_energy} and shown graphically in
Fig.~\ref{fig:Energy_of_form}.

\begin{table}[t!]
\caption{\label{tab:formation_energy} Pauling electronegativity $\chi_p$, calculated energy of formation $\Delta E_\text{atom}$ [eV per atom], and experimentally observed decomposition temperature $T_\text{dec}$ [$^\circ$C] for selected MBAs and their plain MB counterparts, relative to the constituent atoms in their natural states.}
\begin{tabular*}{\columnwidth}{@{}l@{\extracolsep{\fill}}ccr@{}}
\multicolumn{4}{@{}l}{MBAs releasing mostly NH$_3$}           \\\hline
Formula  & $\chi_p$  & $\Delta E_\text{atom}$ & $T_\text{dec}$\\\hline
Sr(BH$_4$)$_2\cdot$4NH$_3$  & 0.95 & $-0.130$ & 14            \\
Sr(BH$_4$)$_2\cdot$2NH$_3$  & 0.95 & $-0.298$ & 130           \\
Sr(BH$_4$)$_2\cdot$NH$_3$   & 0.95 & $-0.432$ & 150           \\
Sr(BH$_4$)$_2$              & 0.95 & $-0.643$ & 400\\
Ca(BH$_4$)$_2\cdot$4NH$_3$  & 1.00 & $-0.309$ & 85            \\
Ca(BH$_4$)$_2\cdot$2NH$_3$  & 1.00 & $-0.286$ & 160           \\
Ca(BH$_4$)$_2\cdot$NH$_3$   & 1.00 & $-0.414$ & 225           \\
Ca(BH$_4$)$_2$              & 1.00 & $-0.622$ & 380 \\
Y(BH$_4$)$_3\cdot$7NH$_3$   & 1.22 & $-0.069$ & 70            \\
Y(BH$_4$)$_3\cdot$6NH$_3$   & 1.22 & $-0.066$ & 80            \\
Y(BH$_4$)$_3\cdot$5NH$_3$   & 1.22 & $-0.097$ & 110           \\
Mn(BH$_4$)$_2\cdot$3NH$_3$  & 1.55 & $-0.016$ & 80            \\[2ex]
\multicolumn{4}{@{}l}{MBAs releasing mostly H$_2$}            \\\hline
Formula  & $\chi_p$ & $\Delta E_\text{atom} $ & $T_\text{dec}$\\\hline
Y(BH$_4$)$_3\cdot$4NH$_3$   & 1.22 & $-0.171$ & 165           \\
Y(BH$_4$)$_3\cdot$2NH$_3$   & 1.22 & $-0.316$ & 190           \\
Y(BH$_4$)$_3\cdot$NH$_3$    & 1.22 & $-0.409$ & 195           \\
Y(BH$_4$)$_3$               & 1.22 & $-0.680$ & 200 \\
Mg(BH$_4$)$_2\cdot$2NH$_3$  & 1.31 & $-0.131$ & 205           \\
Mg(BH$_4$)$_2$              & 1.31 & $-0.339$ & 260\\
Zr(BH$_4$)$_4\cdot$8NH$_3$  & 1.33 & $-0.024$ & 60            \\
Zr(BH$_4$)$_4$              & 1.33 & $-0.423$ & 250\\
Mn(BH$_4$)$_2\cdot$2NH$_3$  & 1.55 & $-0.074$ & 135           \\
Mn(BH$_4$)$_2$              & 1.55 & $-0.238$ & 155 \\
Al(BH$_4$)$_3\cdot$6NH$_3$  & 1.61 & $-0.057$ & 167           \\
Al(BH$_4$)$_3$              & 1.61 & $-0.345$ & 15 \\
Zn(BH$_4$)$_2\cdot$2NH$_3$  & 1.65 & $-0.087$ & 131           \\
Zn(BH$_4$)$_2$              & 1.65 & $-0.249$ & $-35$\\
NaZn(BH$_4$)$_3\cdot$2NH$_3$  & 0.93, 1.65 & $-0.190$ & 133     \\
Li$_2$Mg(BH$_4$)$_4\cdot$6NH$_3$  & 0.98, 1.31 & $-0.140$ & 80  \\
Li$_2$Al(BH$_4$)$_5\cdot$6NH$_3$  & 0.98, 1.61 & $-0.180$ & 138 \\
\end{tabular*}
\end{table}

The primary result from these data is that all MBAs are
thermodynamically destabilized relative to their plain MB counterparts.
Furthermore, sequential events of NH$_3$ release from the same structure
typically results in a MBA with lower energy of formation per atom. For
instance, Sr(BH$_4$)$_2\cdot$4NH$_3$ may decompose by releasing 2NH$_3$,
leaving Sr(BH$_4$)$_2\cdot$2NH$_3$, which has a lower energy of
formation per atom.

Note that NH$_3$ release in these materials is frequently accompanied by
an endothermic event,\cite{Jepsen_2015:ammine_calcium,
Jepsen_2015:tailoring_properties} whereas lowering energy per atom seems
to point to an exothermic process. However, reaction energies for the
reaction
\begin{equation}
\mathcal{M}{\rm (BH_4)}_x\cdot y{\rm NH_3 \longrightarrow \mathcal{M}(BH_4)}_x\cdot(y-z){\rm NH_3} + z{\rm NH_3}
\end{equation}
are positive, correctly indicating endothermic processes.

The thermodynamic destabilization resulting from ammoniation generally
explains the lowering of decomposition temperatures for MBAs observed in
Fig.~\ref{fig:stability}. However, it does not explain the increase of
decomposition temperatures in high-$\chi_p$ MBAs. As a result, we must
conclude that high-$\chi_p$ MBAs are stabilized kinetically rather than
thermodynamically. We discuss the corresponding pathways and kinetics of
the decomposition processes in the next section.

\subsection{Examining decomposition processes}

The processes by which these MBAs decompose have not been clearly
determined experimentally. The possible pathways range from simple
direct formation of H$_2$ from H$^{\delta +}$ and H$^{\delta -}$ to very
complex interactions forming boron nitrides similar to those found in
the decomposition of NH$_3$BH$_3$. The observed final products are
typically amorphous and poorly characterized. This uncertainty makes the
decomposition process difficult to model; without good guidelines from
experiment the possible search space is vast.

One recent theoretical study\cite{Wang_2016:mechanism_controllable}
simulated the decomposition of LiBH$_4\cdot$NH$_3$ and
Mg(BH$_4$)$_2\cdot$2NH$_3$ and suggests that the decomposition process
begins with the NH$_3$ drifting away from the $\mathcal{M}$(BH$_4$)$_x$,
allowing the $\mathcal{M}$(BH$_4$)$_x$ to polymerize with its neighbors.
Then H$^{\delta +}$ in NH$_3$ attacks H$^{\delta -}$ from [BH$_4$]$^-$.
While those findings are valuable, it is unclear whether they can be
generalized to materials with $\mathcal{M}$ of very different
electronegativities. As mentioned in Section~\ref{sec:dihydrogen}, the
direct combination of H$^{\delta +}$ and H$^{\delta -}$ has been
suggested to be a possible hydrogen production mechanism in low-$\chi_p$
materials,\cite{Jepsen_2015:tailoring_properties} and simulated in a few MBAs,\cite{Chen_2012:electronic_structure} but in high-$\chi_p$
materials, the direct H$_2$ creation pathway would compete with the
B$_2$H$_6$ creation pathway from the plain MB. B$_2$H$_6$ release is
always suppressed in the ammoniated
materials,\cite{Jepsen_2015:trends_syntheses} but it is not immediately
obvious whether this is because the diborane production pathway has been
circumvented or because the produced B$_2$H$_6$ react immediately with
the NH$_3$.  To elucidate these issues, we simulate each of the broad
categories mentioned above---NH$_3$ release, direct H$_2$ formation, and
B$_2$H$_6$ formation---and find that each of them becomes the dominating
pathway for low-, mid-, and high-$\chi_p$ values, respectively.

\subsubsection{NH$_3$ release}

Low-$\chi_p$ materials release NH$_3$ preferentially to H$_2$ or
B$_2$H$_6$ and some mid-$\chi_p$ materials release NH$_3$ as they are
heated before switching to H$_2$ release---see Fig.~\ref{fig:stability}.
In order for an MBA to release NH$_3$, the NH$_3$ must escape the metal
to which it is coordinated and then escape as a gas.

\begin{figure}[t]
\includegraphics[width=\columnwidth]{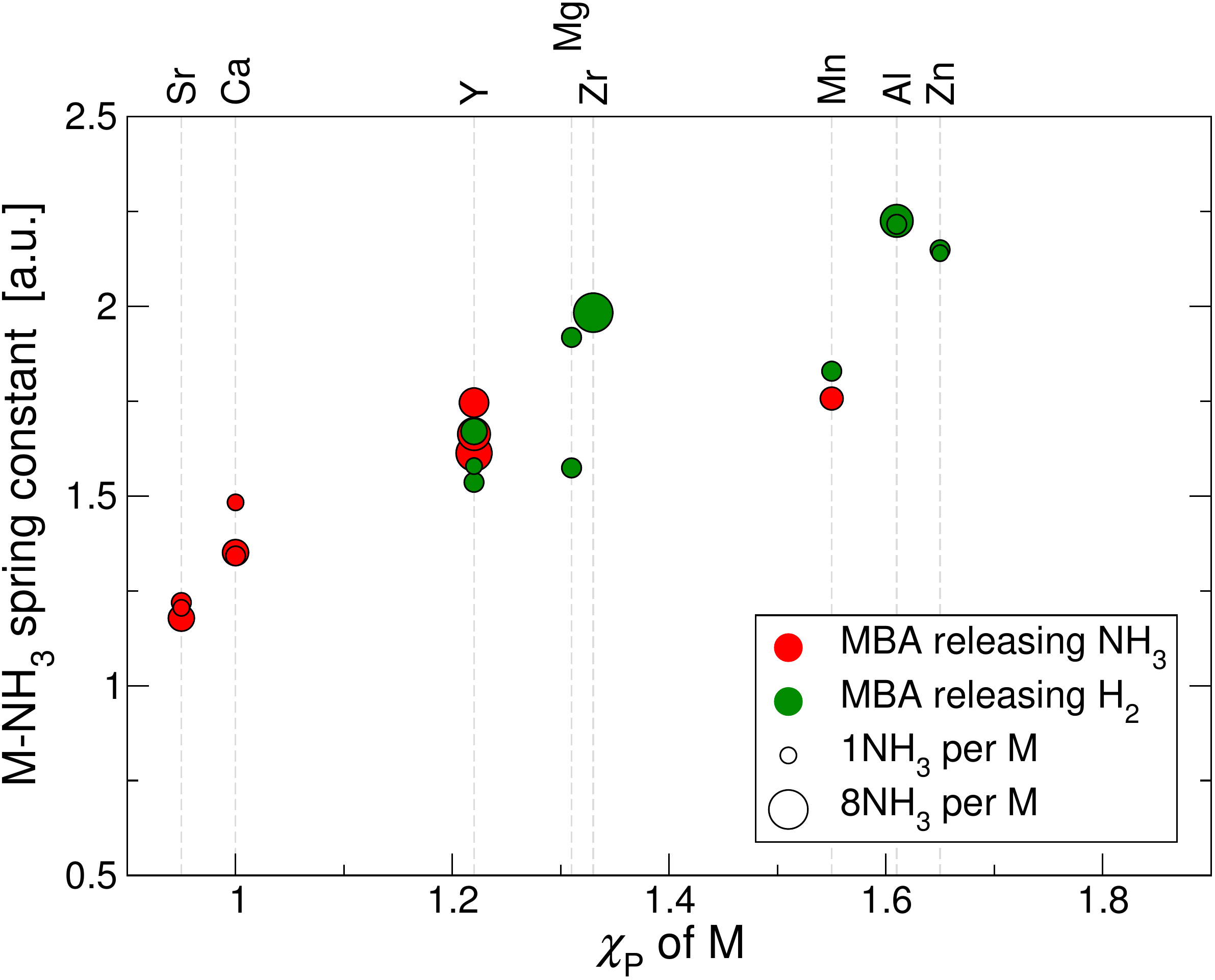}
\caption{\label{fig:MN_stretch} Average spring constants for the $\mathcal{M}$--NH$_3$ bond modeled as a harmonic oscillator, as extracted from short fixed-displacement motion. Red and green spots indicate MBAs that release NH$_3$ and H$_2$ upon decomposition, respectively. The size of the symbol indicates the number of NH$_3$ molecules in the material per metal.}
\end{figure}

We quantitatively studied the behavior of the $\mathcal{M}$-NH$_3$ bond by approximating it as a spring. For small displacements of NH$_3$ with respect to $\mathcal{M}$, we calculated the total energy of the system and based on the increase in energy relative to the ground state we extracted a spring constant from Hooke's law. The
results (depicted in Fig.~\ref{fig:MN_stretch}) show that the NH$_3$
molecules face much steeper potential surfaces in high-$\chi_p$
materials, agreeing with the conventional
wisdom.\cite{Jepsen_2014:boron-nitrogen_based,
Tang_2013:metal_cation-promoted} This means that NH$_3$ in low-$\chi_p$
MBAs can escape more easily, explaining why these materials can
decompose at lower temperatures compared to plain MBs by releasing
NH$_3$, whereas the high-$\chi_p$ MBAs do not---see
Fig.~\ref{fig:stability}. Going from low-$\chi_p$ values to higher ones,
we also see in Fig.~\ref{fig:MN_stretch} that the NH$_3$ release becomes
energetically more and more expensive, suggesting that eventually there
will be a change in decomposition mechanism once an energetically more
favorable pathway becomes available.

In addition to NH$_3$ formation, we must also consider how the NH$_3$ escapes the material before it can be released as a gas. This kind of process may first occur near the surface of the material, creating vacancy defects in the material. Similarly to how Li$^+$ diffuses through LiBH$_4$ as it decomposes by hopping between vacancy defects,\cite{Hoang_2012:mechanism_decomposition} we suggest that NH$_3$ may be able to migrate to the surface through such vacancies. 

\subsubsection{Direct H$_2$ formation}
\label{sec:direct_H2}

\begin{figure}[t]
\includegraphics[width=\columnwidth]{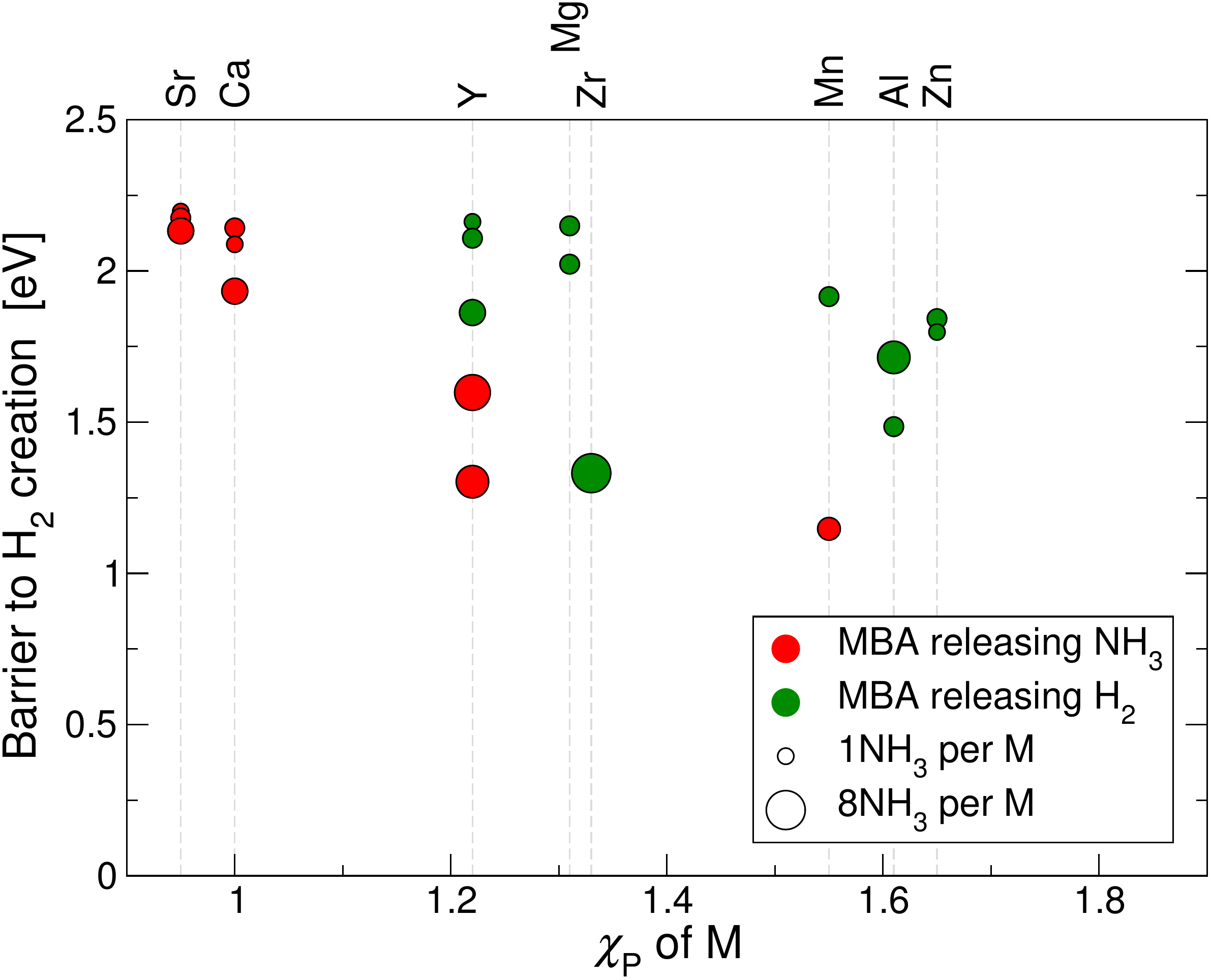}
\caption{\label{fig:H2_stretch} Estimate of transition state energy for the direct H$_2$ formation pathway. Red and green spots indicate MBAs that release NH$_3$ and H$_2$ upon decomposition, respectively. The symbol size indicates the number of NH$_3$ molecules in the material per metal.}
\end{figure}

Direct H$_2$ formation has been suggested as dehydrogenation
pathway\cite{Jepsen_2015:tailoring_properties, Chen_2012:electronic_structure} and can become a more
energetically favorable alternative to NH$_3$ release.  We modeled
direct H$_2$ formation from the H$^{\delta +}$ and H$^{\delta -}$ pairs
that make up the dihydrogen bond network by generating structures that
would be similar to transition states for the reaction. We did this by
moving H$^{\delta +}$ and H$^{\delta -}$ from one dihydrogen bond to a
separation of 0.74~\AA\ along the axis between them and holding their
positions fixed while optimizing the rest of the structure. We did this
relaxation for each dihydrogen bond under 2.5~\AA; the lowest resulting
energies are shown in Fig~\ref{fig:H2_stretch}. NEB calculations
revealed that there is no kinetic barrier between the ground state
structures and the ones that we generated, so we use these energies to
approximate the cost of direct H$_2$ production.

Typically, this procedure resulted in the BH$_3$ shifting to recapture
the H$^{\delta -}$ and the remaining NH$_2$ shortening its distance to
the metal cation, with the net result of a BH$_3$H$_2$, and the H$_2$
would only need to break away from the [BH$_3$]$^-$ and diffuse out of
the material. This result suggests that in an MBA, H is easier to
liberate from NH$_3$ than from [BH$_4$]$^-$ regardless of the metal's
$\chi_p$.

Interestingly, Fig.~\ref{fig:H2_stretch} shows a negative relationship
between $\chi_p$ and the barrier to direct H$_2$ creation; while direct
H$_2$ release is very expensive in low-$\chi_p$ MBAs, it becomes
significantly more favorable for higher-$\chi_p$ materials. One
important note that may not be evident in the plot is that structures
where the metal is directly bound only to NH$_3$ (e.g.
Zr(BH$_4$)$_4\cdot$8NH$_3$ or Al(BH$_4$)$_3\cdot$6NH$_3$) tend to have
lower-energy transition states than the structures where the
$\mathcal{M}$(BH$_4$)$_x$ units are maintained. This may have to do with
BH$_4$ units being less restricted in their movements when they are not
coordinated with any particular metal.

In combination with the previous section we conclude that for
low-$\chi_p$ MBAs the direct H$_2$ release is energetically expensive
(Fig.~\ref{fig:H2_stretch}) whereas NH$_3$ release is favorable
(Fig.~\ref{fig:MN_stretch}). Going to higher $\chi_p$ values, there is a
crossover in those two behaviors and for high-$\chi_p$ we find that
NH$_3$ release is now energetically very expensive, while direct H$_2$
release has become more favorable. This crossover nicely explains the
shift from NH$_3$ release to H$_2$ observed in Fig.~\ref{fig:stability}.
However, it does not explain higher decomposition temperatures after
ammoniation for high-$\chi_p$ materials, which we will further explore
in the last section.

\subsubsection{B$_2$H$_6$ production followed by H$_2$ release}

Up to this point we have found that ammoniation theromodynamically
destabilizes \emph{all} MBs. In addition, the previous two sections help
understand the shift in release product as a function of $\chi_p$.  But,
why do high-$\chi_p$ MBs with $\mathcal{M}$ = Al and Zn get stabilized
upon ammoniation? As we will show, the ammoniation blocks the standard
decomposition pathway that these materials prefer, thus kinetically
stabilizing those MBAs.  We know from the last section that direct H$_2$
release becomes favorable at mid-$\chi_p$ values, but it remains
energetically quite expensive.  Another pathway for MBAs to decompose is
through the formation of B$_2$H$_6$. Previous studies show a strong
correlation between $\chi_p$,\cite{Nakamori_2006:correlation_between}
formation enthalpy,\cite{Harrison_2016:suppressing_diborane} and
decomposition through B$_2$H$_6$ production for MBs. They also show that
this pathway opens up for $\chi_p\gtrsim 1.6$, where it becomes---as we
will show below---an energetically favorable alternative to direct H$_2$
formation.

Jepsen and coworkers note that B$_2$H$_6$ release is always suppressed
in the presence of NH$_3$, where the two react to form ammine metal
borohydrides, suggesting that B$_2$H$_6$ may be an intermediate in the
H$_2$ release pathway.\cite{Jepsen_2015:ammine_calcium} These materials
have a higher decomposition temperature after the ammoniation process,
but at the higher temperature where the MBA begins to decompose, the
same B$_2$H$_6$ formation process may occur. A very low kinetic barrier
of only $\sim$~0.7~eV was found for the
reaction\cite{Nguyen_2008:reactions_diborane}
\begin{equation}\label{equ:B2H6+NH3}
\rm B_2H_6 + NH_3 \longrightarrow NH_2BH_2 + BH_3 + H_2\;,
\end{equation}
making B$_2$H$_6$ formation followed by H$_2$ release energetically
favorable over direct H$_2$ formation.  

If the kinetic barrier to reaction~(\ref{equ:B2H6+NH3}) remains
small in the MBA environment, then the only other possible bottleneck in
MBA decomposition through the B$_2$H$_6$ release pathway is the
formation of the B$_2$H$_6$ intermediate itself.  We performed a
transition state search using the NEB method to find the kinetic
barriers to B$_2$H$_6$ production in selected MBAs, as well as the
materials' plain MB counterparts. To find realistic endpoints, we
simulated perturbing the system in configuration space by pulling nearby
BH$_4$ units together and allowing the system to relax to a local energy
minimum. We performed this process over 250 times and while the
relaxations sometimes yielded the desired B$_2$H$_6$, it more often
resulted in the creation of B$_2$H$_7$---a structure that has been
observed experimentally\cite{Shore_1982:structure_b2h7-} and considered
as a potential metastable intermediate in Mg(BH$_4$)$_2$
decomposition.\cite{Zhang_2012:theoretical_prediction} While B$_2$H$_6$
can lead to hydrogen release via reaction (\ref{equ:B2H6+NH3}) with a
barrier of $\sim0.7$ eV, we found no such reaction with B$_2$H$_7$.
Instead, using NEB calculation we found that the reaction
\begin{equation}\label{equ:NH3+B2H7}
\rm [B_{2}H_{7}]^- + NH_3 \longrightarrow NH_{3}BH_{3} + [BH_{4}]^-
\end{equation}
has a barrier of only $0.75$~eV relative to the NH$_3$ and
[B$_2$H$_7$]$^-$ molecules by themselves in the gas phase---and is thus
equally favorable as reaction (\ref{equ:B2H6+NH3}). After this reaction,
the NH$_3$BH$_3$ is free to decompose on its own (pure NH$_3$BH$_3$
decomposes to NH$_2$BH$_2$, releasing H$_2$ at around 100~$^\circ$C,
facing a kinetic barrier of about 1.5~eV in the gas
phase\cite{Shevlin_2011:dehydrogenation_mechanisms,
Welchman_2015:lowering_hydrogen}) or interact further with [BH$_4$]$^-$
or NH$_3$. In either case, a similar boron nitride may be formed through
the direct H$_2$ formation pathway.

In addition to the low reaction barrier, any produced borane must diffuse to the surface in order to escape the system. While we have not performed diffusion simulations in these materials, we argue that the kinetic barrier for this diffusion should increase in MBAs because the dihydrogen bond network creates a steeper potential surface that the molecules would need to traverse, potentially trapping the molecules in the material until they react with NH$_3$.

\begin{figure}[t]
\includegraphics[width=\columnwidth]{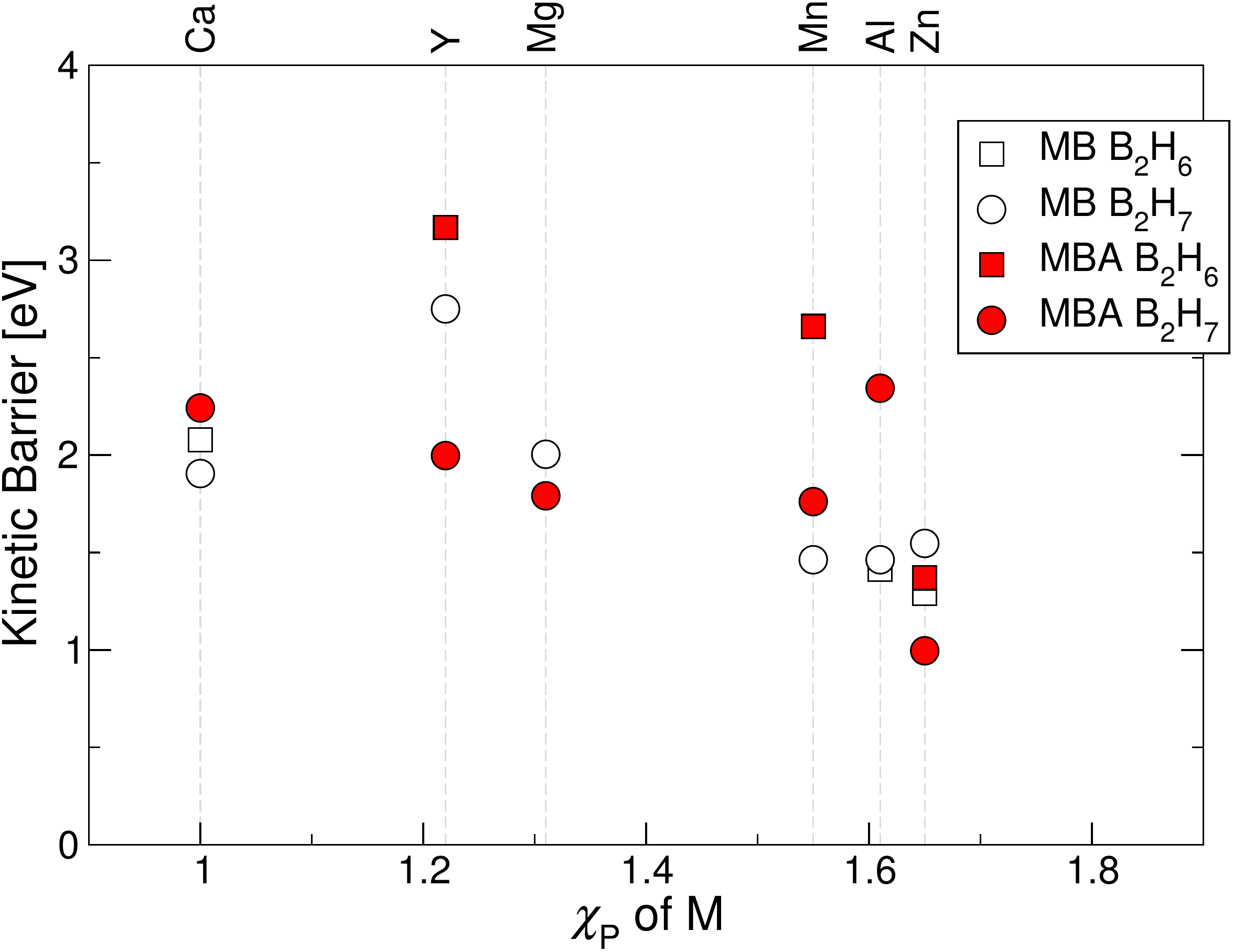}
\caption{\label{fig:barriers} Kinetic barriers to the formation of B$_2$H$_6$ and B$_2$H$_7$ in selected pure MBs and MBAs. Barriers for endpoints that were $\geq3$~eV above the ground state structure were not calculated.}
\end{figure}

So, both B$_2$H$_6$ and B$_2$H$_7$ can eventually produce molecular hydrogen via reactions
(\ref{equ:B2H6+NH3}) and (\ref{equ:NH3+B2H7}) with barriers of only
$\sim 0.7$~eV and the formation of B$_2$H$_6$ and B$_2$H$_7$ themselves
might now be the rate-limiting step. We thus calculated and report
kinetic barriers to the creation of both B$_2$H$_6$ and B$_2$H$_7$ in
Fig.~\ref{fig:barriers}. We did not calculate kinetic barriers to
endpoints that we found were $\geq3$~eV above the ground state
structure.  We find that the kinetic barriers decrease in higher
$\chi_p$ materials, making this pathway viable for $\chi_p\gtrsim 1.6$,
as suggested previously.\cite{Nakamori_2006:correlation_between,
Harrison_2016:suppressing_diborane} In particular, for these
high-$\chi_p$ values, this pathway becomes more favorable than direct
H$_2$ release (compare Figs.~\ref{fig:H2_stretch} and \ref{fig:barriers}).
It only remains to investigate the effect of ammoniation on this pathway
and we analyze the two cases $\mathcal{M}$ = Al and Zn separately.

In Al(BH$_4$)$_3\cdot$6NH$_3$, the ammoniation drastically increases the
barrier for B$_2$H$_7$ formation and we could not even find a pathway
for B$_2$H$_6$ formation.  Ammoniation has thus made this decomposition
pathway practically inaccessible below the material's decomposition
temperature, kinetically stabilizing the Al case.  B$_2$H$_6$ formation
in Zn(BH$_4$)$_2\cdot$2NH$_3$ faces only a slightly increased barrier
upon ammoniation, but a significantly decreased barrier for B$_2$H$_7$
formation, pushing the system towards reaction (\ref{equ:NH3+B2H7}). The
decomposition is then governed by the hydrogen release from
NH$_3$BH$_3$, which occurs around 100
$^\circ$C\cite{Shevlin_2011:dehydrogenation_mechanisms,
Welchman_2015:lowering_hydrogen} and explains the stabilization seen for
Zn in Fig.~\ref{fig:stability}.

The difference between Al and Zn may be due to whether the [BH$_4$]$^-$
anions are able to transfer H$^-$ as they form B$_2$H$_6$. In
Zn(BH$_4$)$_2\cdot$2NH$_3$, the BH$_4$ are still coordinated with the
metal, making the electron transfer simple, but in
Al(BH$_4$)$_3\cdot$6NH$_3$, the NH$_3$ molecules may act as neutral
ligands, preventing reduction of the metal. Roedern and Jensen made a
similar assertion in the case of transition-metal
MBAs.\cite{Roedern_2015:ammine-stabilized_transition-metal}

The reasons differ for Al and Zn, but in both cases, the B$_2$H$_6$
formation pathway that dominates decomposition in the plain MB becomes
unavailable after ammoniation, leading to a higher decomposition
temperature observed in Fig.~\ref{fig:stability}.

\section{Conclusions}

We have examined the decomposition processes of MBAs in order to explain
why the ammoniation process results in lower decomposition temperatures
in materials with low-$\chi_p$ metals and higher ones for those with
high-$\chi_p$ metals.

We found that the dihydrogen bond networks remain very similar between
low and high-$\chi_p$ materials and cannot explain the sharp distinction
between them in terms of the effect of ammoniation.  We also found that
the ammoniation process always \emph{thermodynamically destabilizes}
MBs, while---at the same time---a few high-$\chi_p$ materials get
\emph{kinetically stabilized} through a shift in decomposition
mechanism.  We examined three possible decomposition mechanisms: NH$_3$
release, direct H$_2$ release, and B$_2$H$_6$ or B$_2$H$_7$ formation
followed by H$_2$ release.  At low-$\chi_p$, NH$_3$ is weakly bound to
the metal and the material decomposes by releasing it. Going to
mid-$\chi_p$ values, the release of NH$_3$ becomes less favorable and
the materials switch to direct H$_2$ release, which becomes more
favorable with higher $\chi_p$. Finally, for $\chi_p\gtrsim 1.6$, the
production of B$_2$H$_6$ or B$_2$H$_7$ formation becomes more favorable
than direct H$_2$ release, resulting in indirect H$_2$ release via
case-by-case mechanisms. For the special cases of Al and Zn MBs, we
further found that this pathway becomes inaccessible upon ammoniation
and those phases get kinetically stabilized. In
Al(BH$_4$)$_3\cdot$6NH$_3$, NH$_3$ shields the metal from being reduced,
preventing the electron transfer that occurs in B$_2$H$_6$ formation,
leaving direct H$_2$ formation as the only available decomposition
pathway. In Zn(BH$_4$)$_2\cdot$2NH$_3$, B$_2$H$_7$ formation becomes
more favorable, leading to the formation of NH$_3$BH$_3$ rather than
B$_2$H$_6$.  We can thus explain the peculiar stabilization effect of
ammoniation in MBs and link it to the exact onset of diborane
production.

These insights into how ammoniation affects decomposition processes in
MBs can be used to design new mixed-metal MBA materials that decompose
in a more desirable fashion or suggest better catalysts for
decomposition, perhaps resulting in more reversible hydrogen storage
materials.

\section{Acknowledgements}

This work was supported in full by NSF Grant No.\ DMR--1145968.
T.T. also acknowledges generous support from the Simons Foundation
through Grant No. 391888, which endowed his sabbatical at MIT.

\bibliography{sources} 
\bibliographystyle{rsc}

\end{document}